\documentclass[aps,twocolumn,amsfonts,amsmath,amssymb,mathtools,cuted,prx,superscriptaddress,preprintnumbers,10pt,floatfix,reprint]{revtex4-2}
\usepackage{graphicx,subfigure}
\usepackage{pictexwd,dcpic}
\usepackage{microtype}
\usepackage[hidelinks,colorlinks=true,linkcolor=blue,citecolor=blue]{hyperref}
\usepackage[utf8]{inputenc}
\usepackage{uniinput}
\allowdisplaybreaks

\newcommand{\red}[1]{}
\renewcommand{\red}[1]{{\color{red} {#1}}}
\newcommand{\blue}[1]{}
\renewcommand{\blue}[1]{{\color{blue} {#1}}}

\usepackage{tikz-cd} 
\usetikzlibrary{backgrounds, arrows,calc,shapes,decorations.pathreplacing, decorations.markings, automata,positioning,matrix}

\newcommand\vertarrowbox[3][6ex]{%
  \begin{array}[t]{@{} c c @{}} #2 \\
  \left\downarrow\vcenter{\hrule height #1}\right.\kern-\nulldelimiterspace & #3
  \end{array}%
}

\tikzset{
        cvertex/.style={circle,draw=black,inner sep=1pt,outer sep=3pt},
        vertex/.style={circle,fill=black,inner sep=1pt,outer sep=3pt},
        star/.style={circle,fill=yellow,inner sep=0.75pt,outer sep=0.75pt},
        tvertex/.style={inner sep=1pt,font=\scriptsize},
        gap/.style={inner sep=0.5pt,fill=white}}

\tikzstyle{mybox} = [draw=black, fill=blue!10, very thick,
    rectangle, rounded corners, inner sep=10pt, inner ysep=20pt]
\tikzstyle{boxtitle} =[fill=blue!50, text=white,rectangle,rounded corners]

\usetikzlibrary{arrows}
\tikzstyle{every picture}+=[remember picture]
\tikzstyle{na} = [baseline=-.5ex]
\tikzstyle{mine}= [arrows={angle 90}-{angle 90},thick]

\def\Llleftarrow{%
\lower2pt\hbox{\begingroup
\tikz
\draw[shorten >=0pt,shorten <=0pt] (0,3pt) -- ++(-1em,0) (0,1pt) -- ++(-1em-1pt,0) (0,-1pt) -- ++(-1em-1pt,0) (0,-3pt) -- ++(-1em,0) (-1em+1pt,5pt) to[out=-105,in=45] (-1em-2pt,0) to[out=-45,in=105] (-1em+1pt,-5pt);
\endgroup}
}

\makeatletter
\newcommand\xleftrightarrow[2][]{%
  \ext@arrow 9999{\longleftrightarrowfill@}{#1}{#2}}
\newcommand\longleftrightarrowfill@{%
  \arrowfill@\leftarrow\relbar\rightarrow}
\makeatother

\begin{document}

\title{A new vista on the Heterotic Moduli Space from Six and Three Dimensions}
\author{Michele Del Zotto}
\affiliation{Department of Mathematics,  Uppsala University,  SE-75120 Uppsala, Sweden}
\affiliation{Department of Physics and Astronomy,  Uppsala University,  SE-75120 Uppsala, Sweden}
\author{Marco Fazzi}
\affiliation{Department of Physics and Astronomy,  Uppsala University,  SE-75120 Uppsala, Sweden}
\affiliation{NORDITA, Hannes Alfv\'ens v\"ag 12, SE-10691 Stockholm, Sweden}
\author{Suvendu Giri}
\affiliation{Department of Physics and Astronomy,  Uppsala University,  SE-75120 Uppsala, Sweden}
\affiliation{Department of Physics, Princeton University, Princeton, New Jersey 08544, USA}
\affiliation{Princeton Gravity Initiative, Princeton University, Princeton, New Jersey 08544, USA}

\begin{abstract}
\noindent We settle a long-standing question about the hypermultiplet moduli spaces of the heterotic strings on ALE singularities. These heterotic backgrounds are specified by the singularity type, an instanton number, and a (nontrivial) flat connection at infinity. Building on their interpretation as six-dimensional theories, we determine a class of three-dimensional $\mathcal{N}=4$ quiver gauge theories whose quantum corrected Coulomb branch coincides with the exact heterotic hypermultiplet moduli space.

\end{abstract}
\preprint{UUITP-23/23}
\preprint{NORDITA 2023-109}
\maketitle
\section{Introduction and motivation}

What is the hypermultiplet moduli space of the heterotic string on an ALE singularity? This question was posed long ago by Witten, Sen and many others \footnote{For compactifications on tori, a detailed analysis was carried out in \cite{deBoer:2001wca}.}, and an answer was proposed for singularities of the form $\mathbb C^2/\mathbb Z_k$ both in absence and presence of small instantons \cite{Sen:1997js,Witten:1999fq,Hanany:1999ui}. Upon a further toroidal compactification, this moduli space should be quaternionic K\"ahler \cite{Bagger:1983tt} (i.e. it has holonony contained in $USp(2)\times USp(2n)$ where $n$ is its quaternionic dimension \cite{Besse:1987pua}) and receives $\alpha'$ and worldsheet instanton corrections \cite{Witten:1999eg}. However, in the limit of decoupled gravity it becomes hyperk\"ahler (with $USp(2n)$ holonomy, and a unique Ricci-flat metric) and most corrections disappear. The corrections that remain are such that the resulting space is a smooth manifold, in all calculable examples. For instance, for $\mathbb C^2/\mathbb Z_2$ and in absence of small instantons this moduli space is the celebrated Atiyah--Hitchin manifold $\mathcal{M}_\text{AH}$, of unit quaternionic dimension  \cite{Atiyah:1985fd}. Importantly, this manifold is also the Coulomb branch (CB) of three-dimensional (3D) $\mathcal{N}=4$ pure $SU(2)$ gauge theory, which may be further identified with the moduli space of two BPS 't Hooft--Polyakov monopoles of $SU(2)$. It is then natural to conjecture that the heterotic moduli space on a $\mathbb C^2/\mathbb Z_k$ ALE space should be the CB of pure $SU(k)$, which is the same as the moduli space \footnote{\label{foot:USU}A more precise statement is the following. The \emph{reduced} moduli space of $k$ monopoles of $SU(2)$ (with the center-of-mass modulus factored out) is the same as the CB of pure $SU(k)$, i.e. the $U(1)$ center of $U(k)$ decouples from the dynamics as it is free in the infrared. Alternatively, the moduli space of $k$ $SU(2)$ monopoles is the CB of $SU(k)$ times the moduli space of a free vector.} of $k$ BPS monopoles of $SU(2)$ \cite{Seiberg:1996nz,Hanany:1996ie,Chalmers:1996xh,Dorey:1997ij}. This observation lends itself to a further natural generalization, which led \cite{Witten:1999fq} to propose that the hypermultipet moduli space of the heterotic string on an ALE space locally modeled by a $\mathbb{C}^2/\Gamma_G$ singularity (with $\Gamma_G \subset SU(2)$ finite) should be the same as the CB of 3D $\mathcal{N}=4$ pure $G$ gauge theory, where $G$ is the McKay dual group \cite{McKay} to $\Gamma_G$ \footnote{The pure $G$ theory with standard Yang--Mills kinetic term for the gauge field has $4r$ massless real scalars if $G$ is semisimple of rank $r$. Namely, the quaternionic dimension of its CB is $\dim_\mathbb{H} \text{CB}(G)=r$. The physics is free in the infrared for large vacuum expectation values of the scalars \cite{Seiberg:1996nz}. This can be extended to quivers with ADE shape, $U(n_i)$ gauge groups, and (only) bifundamental matter. The CB has dimension $\dim_\mathbb{H}\text{CB}=\sum_i n_i$ and is interpreted as the moduli space of BPS monopoles of $SU(r+1)$ with $r\geq 1$, if $r$ is the rank of the ADE group \cite{Tong:1998fa}. In type A, this matches what found in \cite{Hanany:1996ie} via Type IIB brane constructions.}.  This conjecture has been verified using a variety of techniques \cite{Rozali:1999va,Mayr:1999bk,Aspinwall:1999xs}. For instance, in the simplest case of $G=SU(k)$, $\Gamma_{SU(k)}=\mathbb Z_k$ and the dimension of the heterotic moduli space is counted by the number of resolution parameters of the $\mathbb{C}^2/\mathbb{Z}_k$ orbifold, i.e. $k-1$. This is indeed the same as the (quaternionic) dimension of the Coulomb branch of pure $SU(k)$.

The above setup requires no instantons of the heterotic gauge group (i.e. no heterotic NS5-branes), that is we take $F=0$ \emph{identically} and thus the instanton number $\text{Tr} F^2$ is also vanishing. The equation of motion for the dilaton in the heterotic string reads schematically $\Delta \phi = \text{Tr} F^2 - \text{Tr} R^2$, with $R$ the Riemann tensor (regarded as a two-form valued in the Lie algebra of the orthogonal group, i.e. the curvature of the spin connection $\omega$). With $F=0$ and $R$ large we are driven to weak string coupling, and we will not see fluctuations around $F=0$ which may distinguish the two heterotic strings. The analysis of \cite{Witten:1999fq} then proceeds by performing a calculation in the heterotic sigma model CFT (valid at weak string coupling) to establish the correct moduli space (the Atiyah--Hitchin manifold in the $k=2$ case of $\Gamma_G = \mathbb{Z}_k$, i.e. $G=SU(2)$).

What happens if we add say $M$ small (i.e. zero-size, or pointlike) instantons of the heterotic string gauge group? On top of the instanton number $M$, these should be specified by the value of a (possibly nontrivial) flat connection $F=0$ \emph{at the spatial infinity} $S^3/\Gamma_G$ surrounding the orbifold point of the ALE space (since $\pi_1(S^3/\Gamma_G) \neq \emptyset$). This flat connection is a representation (i.e. an injective homomorphism, or embedding) of the orbifold group in the gauge group, $\lambda:\Gamma_G \to Spin(32)/\mathbb{Z}_2$ or $\mu:\Gamma_G \to E_8$ (one per $E_8$ factor), for the two possible heterotic strings. These instantons have to be interpreted appropriately in string theory. In the case of the $Spin(32)/\mathbb{Z}_2$ string, they are dual to Type I D5-branes \cite{Witten:1995gx}, whereas in the case of the $E_8 \times E_8$ string they are given by M5-branes in the dual Ho\v{r}ava--Witten M-theory background on an interval \cite{Horava:1995qa,Horava:1996ma}. (More precisely, we will have say $N$ ``full'' instantons given by the M5's plus $N_\mu$ ``fractional'' instantons produced by the fractionalization of each of the two M9-walls against the orbifold \cite{Hanany:1999ui,DelZotto:2014hpa}, so that $M=N+N_\mu$). In the heterotic string, they will act as sources in the Bianchi identity for the strength of the B-field, $dH=\text{Tr} F \wedge F - \text{Tr} R \wedge R$, with $M=\int dH /(4\pi)^2$ in appropriate units \footnote{Equivalently, $H=dB+\text{CS}(A)-\text{CS}(\omega)$, where CS is the Chern--Simons invariant, $A$ the gauge connection, and $\omega$ the spin connection. E.g. $\text{CS}(A)=\text{Tr} (A\wedge dA +\frac{2}{3}A\wedge A \wedge A)$.}.

What is the role of these instantons in the 3D theory? How do they modify its CB so as to capture the new heterotic moduli introduced by them? In the $G=SU(k)$ case it was soon realized \cite{Hanany:1997gh,Rozali:1999va,Aspinwall:1999xs} that their addition in the $E_8 \times E_8$ heterotic string corresponds to adding $M$ flavors (i.e. fundamental hypermultiplets) to the 3D pure $G$ gauge theory, when the flat connection at infinity \emph{is trivial}, i.e. when the full $E_8\times E_8$ group is preserved. Moreover \cite{Hanany:1999ui} gave an interpretation of the $2$ BPS monopoles of $SU(2)$ in the heterotic setup (when $k=2$): they correspond to $2$ half-NS5-branes stuck on an O8$^-$-plane obtained by reducing each of the two $E_8$ M9-walls of the Ho\v{r}ava--Witten setup to Type IIA. (This can be generalized to $k$ NS5s.) 
The question as to what is the 3D theory corresponding to the hypermultiplet moduli space in presence of a \emph{nontrivial} flat connection at infinity was however left open in \cite{Hanany:1999ui}. 

The main question we address in this work is how to capture the quantum corrected hypermultiplet moduli space in presence of a possibly nontrivial flat connection at infinity in the heterotic string. Our main result is that we can still construct 3D theories that capture this space via their CB.
\section{3D quivers from 6D}
Consider the $E_8\times E_8$ heterotic string compactified on a real four-dimensional ALE space, namely a K3 surface with a singularity locally modeled by $\mathbb{C}^2/\Gamma_G$. The resulting 6D theories, dubbed $\mathcal T_M(\mu_\textsc{L},\mu_\textsc{R})$, have recently been completely determined  \cite{DelZotto:2022ohj,DelZotto:2022xrh,dzlo3}, building on \cite{Aspinwall:1997ye,Blum:1997mm,Blum:1997fw,Intriligator:1997dh,Bhardwaj:2015oru}, and have the structure of a fusion \cite{Heckman:2018pqx,DelZotto:2018tcj} of two 6D orbi-instanton superconformal field theories $\Omega_N(\mu)$ \footnote{These are specific 6D theories denoted $\Omega_N(\mu)$, where $\mu : \Gamma_G \to E_8$ is a group homomorphism and $N$ is a non-negative integer (for their classification see e.g. \cite{Heckman:2015bfa,Frey:2018vpw}).}. Schematically:
\begin{equation}
\label{eq:fusion}
\mathcal T_M(\mu_\textsc{L},\mu_\textsc{R}) = \Omega_{N_\textsc{l}}(\mu_\textsc{l}) - G - \Omega_{N_\textsc{r}}(\mu_\textsc{r})\,.
\end{equation}
The Higgs branch of this 6D theory is the quantum corrected hypermultiplet moduli space of the heterotic string $\mathcal M^\text{Het}_{G;M,\mu_\textsc{l},\mu_\textsc{r}}$. Since a fusion is the 6D generalization of a gauging operation, the result is a simple hyperkähler quotient \cite{Hitchin:1986ea}
\begin{equation}\label{eq:quotient6D}
    \mathcal M^\text{Het}_{G;M,\mu_\textsc{l},\mu_\textsc{r}} = (\mathcal M_{\Omega_{N_\textsc{l}}(\mu_\textsc{l})} \times \mathcal M_{\Omega_{N_\textsc{r}}(\mu_\textsc{r})}) \, /\!/\!/ \, G
\end{equation}
of the Higgs branches of the known 6D theories $\Omega_{N_\textsc{l}}(\mu_\textsc{l})$ and $\Omega_{N_\textsc{r}}(\mu_\textsc{r})$. Here, as above,  $M=N_\textsc{l} + N_{\mu_\textsc{l}} + N_\textsc{r} + N_{\mu_\textsc{r}}$. This is the starting point for our paper: the 3D theories of  interest can be determined starting from these 6D models. We will outline the construction in full detail only in type A, i.e. for $G=SU(k)$; the other types are computationally more challenging but not conceptually harder \footnote{Importantly, they may be approached with techniques already available in the literature.}. In each case, the Higgs branch of interest is captured by a 3D $\mathcal{N}=4$ quiver gauge theory, colloquially known as ``magnetic quiver'' \cite{Cabrera:2019izd}, which flows in the infrared (IR) to a 3D $\mathcal{N}=4$ superconformal field theory (SCFT). The hypermultiplet moduli space of the heterotic string is captured by the CB of the moduli space of this SCFT. In the case where we already know the answer, i.e. the case of $M$ small instantons for $G=SU(k)$ with trivial flat connection (i.e. the $E_8 \times E_8$ case), this quiver ought to be the one for $SU(k)$ with $M$ flavors, i.e. 
\begin{equation}
\label{eq:1kCBtrick-pre}
    \overset{\overset{\displaystyle \boxed{M}}{\vert}}{SU(k)}\ .
\end{equation}
This quiver is related to the magnetic quiver construction of  \cite{Cabrera:2019izd} upon replacing $SU(k)$ with $U(k)$ and $\boxed{M}$ with a bouquet of $M$ $U(1)$ gauge nodes. (The opposite operation is the so-called hyperk\"ahler implosion \cite{Dancer:2012sv,Dancer:2020wll}, \footnote{In physics, it simply means we can ungauge a $U(1)$ by gauging the topological $U(1)$ symmetry that comes with it.} preserving the hyperk\"ahler structure of the moduli space and the action of (a maximal torus of) the flavor symmetry group.) The $SU$ group arises in 3D from a reduction of the 6D $SU$ gauge group (since the $U(1)$ centers of unitary groups are massive in 6D, and hence decouple from the low-energy dynamics); the (opposite of the) implosion is related to a gauging of a $S_M$ discrete symmetry which exchanges the $M$ identical M5s/NS5s in 6D (see \cite{Hanany:2018vph} for more details). The result is:
\begin{equation}
\label{eq:1kCBtrick}
\overset{\overset{\displaystyle \overbrace{1 \cdots 1}^{M}}{ \rotatebox[origin=c]{190}{$\setminus$}\ \rotatebox[origin=c]{-190}{$/$} }}{k}
\end{equation}
with an overall $U(1)$ decoupled from the IR dynamics. In \eqref{eq:1kCBtrick-pre} and \eqref{eq:1kCBtrick} (and henceforth) we have adopted standard quiver notation, where a non-negative integer $k$ denotes a 3D $\mathcal{N}=4$ $U(k)$ vector multiplet, an edge connecting two such non-negative integers $k_1$ and $k_2$ denotes a bifundamental hypermultiplet \footnote{I.e., a hypermultiplet transforming in the $(k_1\otimes \overline{k_2}) \oplus (\overline{k_1}\otimes k_2)$ representation of $U(k_1)\times U(k_2)$. \label{foot:bibi}}, and $\boxed{p}-k$ denotes $p$ hypermultiplets transforming in the fundamental of the $U(k)$ gauge group. 

Notice that, for this theory to be ``good'' in the sense of \cite{Gaiotto:2008ak}, i.e. to flow to a standard non-Gaussian (interacting) fixed point in the IR, we \emph{must} have $M \geq 2k$ (or, more generally, $N_\text{f} \geq 2N_\text{c}$) \footnote{This is a consequence of the ``s-rule'' (ensuring that supersymmetry is preserved) in a Type IIB engineering via D3-D5-NS5-branes of this quiver \cite{Hanany:1996ie,Gaiotto:2008ak}}. In other words, a configuration with $M<2k$ does \emph{not} have a mirror dual \cite{Intriligator:1996ex}. For this reason, one \emph{cannot} simply take the $M=0$ limit and get back to Witten's original configuration with no heterotic instantons. 

More generally, we will construct 3D quivers whose CB captures the hypermultiplet moduli space of the $E_8\times E_8$ heterotic string on $\mathbb{C}^2/\mathbb{Z}_k$ for any $k$ and choice of flat connection at infinity. The latter (trivial and nontrivial alike) are classified by group homomorphisms, i.e. embeddings, $\mu_\textsc{L,R}: \mathbb{Z}_k \to E_8$ (one per gauge group factor). Each such embedding is specified by a so-called Kac label \cite{kac1990infinite}, i.e. an integer partition of $k$ in terms of the Coxeter labels $1,\ldots,6,4',3',2'$ of the \emph{affine} $E_8$ Dynkin diagram:
\begin{equation}
\label{eq:kac}
    k=\left(\sum_{i=1}^6 i n_i\right) + 4 n_{4'} + 3 n_{3'} +2 n_{2'}\ ,
\end{equation}
which will be denoted in general $\mu_\textsc{L,R}=[1^{n_1},\ldots,6^{n_6},4^{n_{4'}},3^{n_{3'}},2^{n_{2'}}]$ (and we will also say that the $n_i,n_{i'}$ are the multiplicities of the parts of the Kac label). Each embedding preserves a subalgebra of $E_8$ determined via a simple algorithm: one simply ``deletes'' all nodes with nonzero multiplicity $n_i,n_{i'}$ in this partition, and reads off the Dynkin of the leftover algebra, which may be a sum of nonabelian algebras plus a bunch of $\mathfrak{u}(1)$'s. As an example, the trivial flat connection (embedding), which exists for any $k$, is given by the label $k=[1^k]$, and preserves the full $E_8$. In this case, as stated above, the moduli space is given by the CB of \eqref{eq:1kCBtrick}. As an example of nontrivial flat connection, consider the case $k$ even, which admits a partition $k=[2^m]$ and corresponds to a nontrivial flat connection breaking the heterotic gauge group to $E_7 \times SU(2)$.

Can we see a more direct engineering of these 3D quivers from the heterotic string? The answer is a resounding \emph{yes}, and comes from looking at the compactification of the heterotic string on a K3 as a so-called 6D $(1,0)$ little string theory (LST) \cite{Seiberg:1997zk}. In the $E_8 \times E_8$ case in presence of $M$ small instantons, with nontrivial flat connection at infinity, such LSTs are precisely the theories $\mathcal T_M(\mu_\textsc{L},\mu_\textsc{R})$ in equation \eqref{eq:fusion}. The latter can be given a quiver description via ``geometric engineering'' in terms of F-theory \cite{Aspinwall:1997ye}, which we reproduce below in \eqref{eq:little2} for the case $\mu_{\textsc{l}}=\mu_{\textsc{r}}=[1^k]$. Taking the $T^3$ compactification of this quiver produces a 3D theory which we will call ``electric quiver'', and which reads
\begin{widetext}
\begin{equation}
\label{eq:first3D}
    1 - 2 - 3 - \cdots - (k-1) - \underbrace{\underset{\overset{\displaystyle \vert}{\displaystyle 1}}{k} - k - \cdots - k - \underset{\overset{\displaystyle \vert}{\displaystyle 1}}{k}}_{M-2k+1} - (k-1) - \cdots - 3 -2 -1\ ,
\end{equation}
\end{widetext}
with an overall $U(1)$ decoupled from the IR dynamics \cite{Hanany:1996ie}. The above quiver is \emph{mirror dual} to the one in \eqref{eq:1kCBtrick-pre} \cite{Hanany:1996ie,Intriligator:1996ex}: it clearly shows that we must have $M-2k+1 \geq 1$, i.e. $M \geq 2k$ (which is the same as the s-rule constraint valid in \eqref{eq:1kCBtrick}), ensuring the existence of a mirror in the first place. 

The rest of the paper showcases general 3D quivers (derived in the companion paper \cite{DelZotto:2023nrb}), whose CB captures the hypermultiplet moduli space of interest in the heterotic string. As outlined above, when the latter is compactified on a K3, and moreover gravity is decoupled, its dynamics is captured by a 6D little string theory of type $\mathcal T_M(\mu_\textsc{L},\mu_\textsc{R})$, which in turn is obtained by ``gluing'' together two 6D SCFTs known as orbi-instantons as in equation \eqref{eq:fusion}. We will consider LSTs specified by $M$ small instantons, where
\begin{equation}
\label{eq:Ms}
M=M_\textsc{l}+M_\textsc{r}=(N_\textsc{l}+N_{\mu_\textsc{l}})+(N_\textsc{r}+N_{\mu_\textsc{r}})\ .
\end{equation}
$N \equiv N_\textsc{l}+N_\textsc{r}$ is the total number of \emph{full} heterotic NS5's and $N_{\mu_\textsc{l,r}}$ is the number of fractional instantons in the left and right ``half'' (orbi-instanton) of the LST. This piece of data can be defined in terms of $\mu_\textsc{L,R}$, generalizing appropriately the case $N_{\mu_\textsc{l,r}}=k$ of a trivial flat connection.

The ``minimal'' case (i.e. no full instantons) has $N=0$; however, as is clear from the above formula, this does \emph{not} imply that $M=0$. This is because each M9-wall in the Ho\v{r}ava--Witten setup fractionates once we introduce the orbifold $\mathbb{C}^2/\mathbb{Z}_k$ \cite{DelZotto:2014hpa}. The number of fractions $N_{\mu_\textsc{l,r}}$ (corresponding to \emph{new} NS5-branes in the dual Type I' setup) depends on the specific choice of embedding $\mu_\textsc{L,R}$. For instance, in \eqref{eq:1kCBtrick}-\eqref{eq:first3D} we made the choice $\mu_\textsc{L,R}=[1^k]$ implying $M=2k$ if $N=0$. In this case, we say the magnetic quiver is exactly balanced (since $N_\text{f}=2N_\text{c}$) \cite{Gaiotto:2008ak}. It may happen that for other choices of $\mu_\textsc{L,R}$ the quiver is overbalanced (i.e. $N_\text{f}>2N_\text{c}$). Importantly, we will show that the quiver is never underbalanced (i.e. bad, in the sense of \cite{Gaiotto:2008ak}) for any choice of $\mu_\textsc{L,R}$.

\smallskip 

\section{Coulomb branch and heterotic moduli spaces}

The upshot of our companion paper \cite{DelZotto:2023nrb} is the construction of new 3D $\mathcal{N}=4$ QFTs flowing in the IR to SCFTs, whose CB captures the hypermultiplet moduli space of the heterotic string on the orbifold with a choice of flat connection at infinity. Such QFTs are encoded in the following quiver diagrams:
\begin{widetext}
\begin{equation}
       \Bigg(M_\text{L}{E_8^{(1)}}^\vee +\hspace{0.1cm} {\scriptstyle r_{2'}^\text{L} - r_{4'}^\text{L} - \overset{\underset{\scriptscriptstyle \vert}{\scriptstyle r_{3'}^\text{L}}}{r_6^\text{L}}-r_5^\text{L}-r_4^\text{L}-r_3^\text{L}-r_2^\text{L}-r_1^\text{L} }
               \Bigg) - k - \Bigg({\scriptstyle r_1^\text{R}-r_2^\text{R}-r_3^\text{R}-r_4^\text{R}-r_5^\text{R}-\overset{\underset{\scriptscriptstyle \vert}{\scriptstyle r_{3'}^\text{R}}}{r_6^\text{R}}-r_{4'}^\text{R}-r_{2'}^\text{R}} \hspace{0.1cm}
               + M_\text{R}E_8^{(1)} 
               \Bigg)\ ,
         \label{eq:generalinf}
\end{equation}
\end{widetext}
where in the formula above the sums in the parentheses are performed node-by-node and the ranks $r_i^\textsc{l,r}, r_{i'}^\textsc{l,r}$ of the $U$ gauge groups along the tails (some of which may be zero) are determined by the specific Kac label chosen to determine the embeddings $\mu_\textsc{l}$ and $\mu_{\textsc{r}}$. The explicit algorithm was obtained in \cite{Cabrera:2019dob} and is summarized below. Moreover $E_8^{(1)}$ stands for the quiver of \emph{affine} $E_8$ Dynkin shape, with the ranks of the $U$ groups appearing therein being equal to the Coxeter labels, and ${E_8^{(1)}}^\vee$ is the Dynkin mirrored around the vertical axis, i.e. with the bifurcated tail on the left:
\begin{align}
    E_8^{(1)} &= 1-2-3-4-5-\overset{\underset{\scriptstyle \vert}{\displaystyle 3}}{6}-4-2\ , \\
    {E_8^{(1)}}^\vee &= 2-4-\overset{\underset{\scriptstyle \vert}{\displaystyle 3}}{6}-5-4-3-2-1\ .
\end{align}
It is easy to convince oneself that the above quiver is generically overbalanced, sometimes balanced, but never underbalanced, and hence it always flows to a well-defined 3D $\mathcal N=4$ SCFT in the IR. We also have
\begin{widetext}
    \begin{equation} \label{eq:finaldim}
    \dim_\mathbb{H} \mathcal M^\text{Het}_{G;M,\mu_\textsc{l},\mu_\textsc{r}} = 30M +k-1 + \left(\sum_{i=1}^6 r_i^\text{L}\right)+r_{4'}^\text{L}+r_{3'}^\text{L}+r_{2'}^\text{L}+\left(\sum_{i=1}^6 r_i^\text{R}\right)+r_{4'}^\text{R}+r_{3'}^\text{R}+r_{2'}^\text{R}\ ,
\end{equation}
\end{widetext}
with
\begin{equation}\label{eq:rj}
    r_j = (1-\delta_{j6})\sum_{i=1}^{6-j} i n_{i+j} + 2n_{2'}+3n_{3'}+4n_{4'}
\end{equation}
for $j=1,\ldots,6$, and
\begin{subequations}\label{eq:rjprime}
\begin{align}
& r_{2'}= n_{3'}+n_{4'}\ , \\
& r_{3'} = n_{2'}+n_{3'}+2n_{4'}\ , \\
& r_{4'} = n_{2'}+2n_{3'}+2n_{4'}\ .
\end{align}
\end{subequations}
The dimension in \eqref{eq:finaldim} is obtained by summing the dimensions $\dim_\mathbb{H}\text{CB}_\text{3D}$ of the CBs of the quivers in the parentheses in \eqref{eq:generalinf} and subtracting $\dim_{\mathbb{R}}SU(k)=k^2-1$ because of the hyperk\"ahler quotient by $SU(k)$ performed to glue the magnetic quivers of left and right orbi-instanton according to \eqref{eq:quotient6D}. It reduces to what we have already computed below \eqref{eq:first3D} in the case $\mu_\textsc{l}=\mu_\textsc{R}=[1^k]$, for which $r_i=r_{i'}=0$ for all $i$ (left and right), according to the following observation. In that case, \eqref{eq:generalinf} is nothing but the ``infinite-coupling'' magnetic quiver obtained from \eqref{eq:1kCBtrick} performing a total of $M$ small $E_8$ instanton transitions, which turn the bouquet of $M$ 1's into a sum of $M$ $E_8^{(1)}$ tails, or more precisely to $M_\textsc{l}$ ${E_8^{(1)}}^\vee$ tails to the left, and $M_\textsc{r}$ $E_8^{(1)}$ to the right of node $k$ there. The dimension of its CB jumps by $29M$, going from $M+k-1$ to $30M+k-1$. More general checks can be found in \cite{DelZotto:2023nrb}.

As a simple consistency check of the above we can reproduce the case with no full small instantons (closer to the original setup by Witten and Sen); it simply amounts to taking $N=0$ in \eqref{eq:Ms} so that we are left only with the ``inevitable'' fractional instantons coming from the fractionalization of each of the M9's (left and right) against the orbifold. It is easy to determine explicitly $N_{\mu_{\textsc{l,r}}}$ from  F-theory via a case-by-base analysis \cite{DelZotto:2022ohj,DelZotto:2022xrh,dzlo3}. In the case of singularities of type A, however, an explicit algorithm determining $N_{\mu_\textsc{l,r}}$ from $\mu_\textsc{L,R}$ was found in \cite{Fazzi:2023ulb}. Once one constructs the Type I' setup dual to the heterotic string on the orbifold one can easily read off, for each choice of $(\mu_\textsc{l},\mu_\textsc{r})$, the so-called largest linking number $l_\text{L,R}$ on the left and right of the setup, i.e. the largest among the numbers $(l_1^\text{L,R},\ldots,l_8^\text{L,R},l_9^\text{L,R})$, which are defined as the number of D6's ending from the right on the $i$-th D8, minus the number of D6's ending on the left, plus the number of NS5's to the immediate left of it (where for concreteness we have assumed the O8 sits on the left of each half, when considered individually). More precisely, we have the relation
\begin{equation}
    \label{eq:Nl}
     N_{\mu_\textsc{l,r}} = l_\text{L,R} = \sum_{i=1}^6 n_i + p
 \end{equation}
 with $p=\min \left( \lfloor \left(n_{3'}+n_{4'}\right)/2 \rfloor, \lfloor   \left(n_{2'}+n_{3'}+2n_{4'}\right)/3 \rfloor \right)$.
To determine the Type I' configuration dual to the heterotic string, or equivalently to find the electric quiver in the F-theory notation, one can apply the algorithm of \cite{Mekareeya:2017jgc}. For instance, for $\mu_\textsc{l}=\mu_\textsc{r}=[1^k]$ it produces
\begin{widetext}
\begin{equation}
\label{eq:little2}
[E_8] \, \underbrace{\overset{\emptyset}{1}\, \overset{\mathfrak{su}(1)}{2} \overset{\mathfrak{su}(2)}{2} \cdots \overset{\mathfrak{su}(k-1)}{2}}_{N_{\mu_\textsc{l}}=k} \underbrace{\overset{\mathfrak{su}(k)}{\underset{[N_\text{f}=1]}{2}} \overset{\mathfrak{su}(k)}{2}\cdots \overset{\mathfrak{su}(k)}{2} \overset{\mathfrak{su}(k)}{\underset{[N_\text{f}=1]}{2}}}_{N_\textsc{l}+N_\textsc{r}+1=N+1} \underbrace{\overset{\mathfrak{su}(k-1)}{2} \cdots \overset{\mathfrak{su}(2)}{2} \overset{\mathfrak{su}(1)}{2} \, \overset{\emptyset}{1}}_{N_{\mu_\textsc{r}}=k} \,[E_8] \ .
\end{equation}
\end{widetext}
(All more general choices have been analyzed in detail in \cite{Bhardwaj:2015oru,DelZotto:2022ohj,DelZotto:2022xrh,dzlo3}.) The $T^3$ compactification of the above electric quiver is nothing but \eqref{eq:first3D}, which is mirror to \eqref{eq:1kCBtrick}. The dimension of the HB of the above quiver equals $M+k-1$, as expected. To obtain $30M+k-1$ we first move to infinite coupling in each of the two orbi-instanton constituents, compute the dimension of their HB, and subtract the (real) dimension of the diagonal $SU(k)$ flavor group from the two which we are gauging, producing one exra $\mathfrak{su}(k)$ algebra \footnote{Remember that the quaternionic dimension of the HB in a theory with eight supercharges in any spacetime dimension is found by subtracting the total number of vector multiplets from that of hypermultiplets.}.

\section{Conclusions}

In this paper we have identified the hypermultiplet moduli space of the $E_8 \times E_8$ heterotic string on an A-type ALE space with the exact quantum corrected CB of a 3D quiver gauge theory (the magnetic quiver \eqref{eq:generalinf}) flowing in the IR to an SCFT. The magnetic quiver is the mirror of the $T^3$ compactification of a 6D quiver gauge theory (plus tensor multiplets) that gives the dynamics of a 6D (1,0) LST with a Higgs branch that coincides with the heterotic moduli space (in the limit of decoupled gravity). To obtain this magnetic quiver, one also has to perform $M$ small $E_8$ instanton transitions in 6D.

The dimension of the moduli space in question can be computed both in absence and in presence of $N$ full small instantons, where one is also required to specify a flat connection at infinity for the heterotic gauge group. The various choices of flat connection give rise to different magnetic quivers, with different CBs.
One is moreover led to conjecture (due the same string theory arguments used by \cite{Hanany:1999ui}) that this CB (which is a hyperk\"ahler space) is smooth, as was the case in absence of small instantons \cite{Sen:1997js,Witten:1999fq}. It remains to be understood what is the underlying geometry of the CB. In the companion paper \cite{DelZotto:2023nrb}, based on observations about 6D $(1,0)$ SCFTs made in \cite{Fazzi:2022hal,Fazzi:2022yca,Fazzi:2023ulb}, we have proposed that this CB is given by the holomorphic symplectic quotient construction of \cite{Moore:2011ee} applied to two strata of the so-called affine Grassmannian of $E_8$. It would be extremely interesting to prove this conjecture. The minimum symmetry on this CB in the IR can also be easily computed, which may help with the proof.

As a final remark, the hyperkähler/quaternionic Kähler correspondence \cite{Cecotti:1988qn,Hitchin:2012vvn} suggests that one can also determine  the gravitational quantum moduli space of the heterotic string on these backgrounds building from our results.

\bigskip

\section*{Acknowledgments}

MDZ would like to thank Timo Weigand for an interesting conversation at the Strings and Geometry 2023 conference held at the University of Pennsylvania which sparked our interest into this subject, and Kantaro Ohmori for several enlightening discussions. We would like to thank Sav Sethi for interesting correspondence. MF would like to thank the University of Milano--Bicocca, the Pollica Physics Center, the University of Lancaster for hospitality during the completion of this work. MF and SG gratefully acknowledge support from the Simons Center for Geometry and Physics, Stony Brook University (2023 Simons Physics Summer Workshop) at which some of the research for this paper was performed. The work of MDZ and MF has received funding from the European Research Council (ERC) under the European Union's Horizon 2020 research and innovation program (grant agreement No. 851931). MDZ also acknowledges support from the Simons Foundation Grant \#888984 (Simons Collaboration on Global Categorical Symmetries). The work of MF is also supported in part by the Knut and Alice Wallenberg Foundation under grant KAW 2021.0170, the Swedish Research Council grant VR 2018-04438, the Olle Engkvists Stiftelse grant No. 2180108. The work of SG was conducted with funding awarded by the Swedish Research Council grant VR 2022-06157.

\bibliography{main}
\bibliographystyle{apsrev4-1}

\end{document}